\documentclass[preprint,11pt]{elsarticle}

\usepackage{graphicx}
\usepackage{booktabs}
\usepackage{amsmath}
\usepackage{amssymb}
\usepackage{url}
\usepackage{xcolor}
\usepackage{listings}
\usepackage{multirow}
\usepackage{array}
\usepackage{pifont}
\usepackage[hidelinks]{hyperref}

\DeclareUnicodeCharacter{2460}{\ding{192}} % ①
\DeclareUnicodeCharacter{2461}{\ding{193}} % ②
\DeclareUnicodeCharacter{2462}{\ding{194}} % ③

\definecolor{codegray}{gray}{0.95}
\lstdefinestyle{diff3}{
  basicstyle=\ttfamily\footnotesize,
  backgroundcolor=\color{codegray},
  breaklines=true,
  columns=fullflexible,
  keepspaces=true,
  frame=single,
  framesep=4pt,
  xleftmargin=6pt,
  showstringspaces=false
}

\journal{arXiv preprint}

\begin{document}

\begin{frontmatter}

\title{Can Large Language Models Resolve Real Java Merge Conflicts?\\
An Evaluation with a Calibrated LLM-as-Judge}

\author[vt]{Bowen Shen}
\ead{bowenshe@vt.edu}
\affiliation[vt]{organization={Virginia Polytechnic Institute and State University},
  city={Blacksburg},
  state={VA},
  postcode={24060},
  country={USA}}

\begin{abstract}
Merge conflicts are a recurring cost of collaborative software development, and the
traditional structured and semi-structured merge tools that address them frequently
\emph{abstain}: when their heuristics do not apply, they leave the conflict unresolved.
Large language models (LLMs) can instead produce a candidate resolution for essentially
any conflict, but measuring whether those resolutions are actually \emph{good} at scale is
itself hard, because obtaining human desirability judgments for every model output does not
scale. This paper studies both problems together on real Java merge conflicts drawn from
ConflictBench. We first build an LLM solver as a \emph{generate--validate--retry} agent that
uses only inference-time signals (conflict markers, a Java parser, and duplicate-declaration
checks) and never sees the developer's answer. We then evaluate its resolutions with a
two-metric suite: (1)~a \emph{developer-match} LLM-as-judge implemented as a G-Eval metric and,
crucially, \emph{calibrated against ConflictBench's human labels before being used}, and
(2)~a deterministic \emph{structural-validity} check that does not use an LLM. On a
meta-evaluation of $n{=}292$ human-labeled cases, the judge achieves \textbf{100\% precision
(zero false accepts)} at 64.6\% recall, so every acceptance it emits is trustworthy and every
downstream rate is a conservative lower bound. Under this validated judge, LLM solvers match the
developer's own resolution on \textbf{$\approx$55\%} of true conflicts (conservative floor), and
under a coverage-fair comparison the LLMs (55--59\%) beat the strongest traditional tool
(AutoMerge, 36.7\%) by roughly 18--22 points---an edge that comes almost entirely from
\emph{coverage}, not raw accuracy: the tools abstain on 20--90\% of conflicts while the LLM under
forced resolution abstains on none. Finally, we find the LLM judge accepted 4 of the 5 resolutions
that fail the deterministic structural check, concrete evidence that structural correctness must
\emph{not} be delegated to an LLM. We release the harness as an open engineering artifact.
\end{abstract}

\begin{highlights}
\item We evaluate modern LLMs on real Java merge conflicts from ConflictBench using a two-layer
harness that strictly separates a no-ground-truth solver from a ground-truth evaluator.
\item The developer-match LLM-as-judge is validated against human labels \emph{first}
(precision 100\%, recall 64.6\%, $n{=}292$), so every reported rate is a conservative lower bound.
\item Under the validated judge, LLMs match the developer resolution on $\approx$55\% of true
conflicts and beat the strongest traditional tool by 18--22 points under a coverage-fair convention.
\item The LLM's advantage is coverage, not accuracy: structured tools abstain on 20--90\% of
conflicts; the LLM under forced resolution abstains on none.
\item A deterministic structural check catches invalid code the LLM judge waved through (① accepted
4 of 5 structural failures), showing structural correctness must not be judged by an LLM.
\end{highlights}


\end{frontmatter}

% ============================================================
\section{Introduction}
\label{sec:intro}

In collaborative software development, programmers create branches to add features and fix bugs
in parallel, and later merge those branches to integrate their edits. When two branches edit the
same region of a file in divergent ways, the edits \emph{conflict} and cannot be co-applied
automatically. Resolving these conflicts is a well-documented drain on developer time, and a large
body of work has built structured and semi-structured merge tools to detect and resolve them
automatically~\cite{apel2011semistructured,apel2012structured,shen2019intellimerge,zhu2018automerge}.

A defining property of these traditional tools is that they \emph{abstain} whenever their
heuristics do not apply. A structured merge tool that cannot align two abstract syntax trees, or a
line-based tool that sees genuinely overlapping edits, simply leaves the conflict markers in place
and reports an unresolved conflict. Abstention is safe, but it means the tool contributes nothing on
exactly the cases a developer most wants help with. Our earlier benchmark study, ConflictBench,
quantified this: across five state-of-the-art tools, applicability varied widely and no tool
resolved more than a fraction of real conflicts~\cite{shen2024conflictbench}.

Large language models (LLMs) change the shape of the problem. An LLM prompted with a conflict will
almost always \emph{produce} a candidate resolution, whether or not the conflict is ``easy.'' This
raises two questions that this paper addresses together:

\begin{itemize}
\item[\textbf{Q1.}] \textbf{Can we trust an automatic judge?} Human desirability judgments do not
scale to every model output, so we need an automatic evaluator. But an LLM-as-judge is only useful
if we know how far to trust it. \emph{How accurately does a calibrated LLM-as-judge reproduce human
desirability labels on merge-conflict resolutions?}
\item[\textbf{Q2.}] \textbf{How good are LLM resolutions?} \emph{Under that validated judge, how
often do modern LLMs match the developer's own resolution on real Java conflicts, and how does that
compare to traditional merge tools once we account for abstention?}
\end{itemize}

We answer these on real Java conflicts from ConflictBench with a harness we call
\emph{ConflictAgent}. Our design rests on a strict two-layer separation
(Fig.~\ref{fig:dataflow}). The \emph{solver layer} is an agent that generates a candidate,
validates it with inference-time signals only, and retries on failure---it never sees the
developer's answer or any judge verdict. The \emph{evaluation layer} runs afterward and is allowed
to use ground truth: it compares the candidate against the developer's actual resolution using two
orthogonal metrics.

The key methodological move is that we \emph{calibrate the judge before we use it}. We first treat
ConflictBench's recorded tool outputs and their human desirability labels as ground truth, ask the
judge to grade those same tool outputs, and measure how well it reproduces the human labels. Only
after the judge passes this meta-evaluation do we use it to grade LLM resolutions. On $n{=}292$
human-labeled cases the judge achieves \textbf{100\% precision} (zero false accepts) at 64.6\%
recall. The zero-false-accept property is what makes the rest of the study interpretable: when the
judge says a resolution is acceptable, it is trustworthy; the cost is that the judge under-credits
acceptable alternatives, so every rate we report is a \emph{conservative lower bound}.

\paragraph{Contributions.} This paper makes the following contributions:
\begin{enumerate}
\item A two-layer evaluation harness for LLM merge-conflict resolution that strictly separates a
no-ground-truth solver agent from a ground-truth evaluator (Sec.~\ref{sec:method}).
\item A developer-match LLM-as-judge that is \emph{validated against human labels before use}
(precision 100\%, recall 64.6\%, $n{=}292$), turning every downstream number into a defensible
lower bound (Sec.~\ref{sec:judge}).
\item An evaluation showing that LLMs match the developer resolution on $\approx$55\% of true
conflicts and, under a coverage-fair convention, beat the strongest traditional tool by 18--22
points---an edge driven by coverage rather than accuracy (Sec.~\ref{sec:results}).
\item Evidence that structural correctness must be judged deterministically: the LLM judge accepted
4 of the 5 resolutions that fail a deterministic parse/structure check (Sec.~\ref{sec:results-struct}).
\end{enumerate}

We frame this work as an \emph{evaluation study} built on an open engineering artifact, not as a new
benchmark: the data, developer resolutions, human labels, and five tool baselines all come from
ConflictBench~\cite{shen2024conflictbench}; our contribution is the LLM agent and the calibrated
evaluation methodology around it.

% ============================================================
\section{Background and Related Work}
\label{sec:background}

\subsection{Three-way merge and conflicts}
Version-control systems merge branches with a three-way \texttt{diff3} algorithm that compares each
branch against the common ancestor (\emph{base})~\cite{khanna2007diff3}. When both branches change
overlapping lines, \texttt{diff3} emits conflict markers (\texttt{<<<<<<<}, \texttt{|||||||},
\texttt{=======}, \texttt{>>>>>>>}) delimiting the left, base, and right variants. Following
ConflictBench, we distinguish \emph{true} conflicts (the two sides make genuinely incompatible
changes that require human judgment) from \emph{false} conflicts (textually overlapping but logically
compatible edits that a good tool should merge automatically)~\cite{shen2024conflictbench}. A prior
characterization study found that developers resolve a large fraction of true textual conflicts by
keeping one branch's edits wholesale~\cite{shen2023characterization}, which is part of why simple
side-picking baselines are a meaningful comparison point (Sec.~\ref{sec:eval}).

\subsection{Automatic merge tools}
Line-based tools such as KDiff3~\cite{kdiff3} operate purely on text.
Semi-structured and structured merge tools---FSTMerge~\cite{apel2011semistructured},
JDime~\cite{apel2012structured}, IntelliMerge~\cite{shen2019intellimerge}, and
AutoMerge~\cite{zhu2018automerge}---exploit the syntax and, in some cases, the refactoring history
of the program to merge more precisely. Their common limitation, quantified in
ConflictBench~\cite{shen2024conflictbench}, is coverage: when a tool's structural assumptions do not
hold it abstains, leaving the conflict unresolved. We reuse exactly these five tools and their
recorded outputs as baselines.

\subsection{LLMs for code and for merge resolution}
LLMs have become strong at code generation and program repair~\cite{chen2021codex,jimenez2024swebench},
and prior learning-based work has targeted merge conflicts directly; MergeBERT, for example, framed
resolution as a classification over a fixed set of edit patterns~\cite{svyatkovskiy2022mergebert}.
Our focus is different: rather than train a model, we evaluate how well modern general-purpose LLMs
resolve real conflicts under an honest, calibrated measurement protocol, and we compare them to
traditional tools on the same judged scale.

\subsection{LLM-as-a-judge}
Using an LLM to evaluate open-ended outputs is now common~\cite{zheng2023judge}, and G-Eval showed
that a chain-of-thought prompt with a form-filled, probability-weighted score aligns better with
humans than a single binary verdict~\cite{liu2023geval}. The recognized risks are self-preference
(a model favoring its own family's outputs) and miscalibration. We address the first by making the
judge a \emph{different vendor} from both solvers, and the second by \emph{validating the judge
against human labels before using it}---the step that distinguishes a trustworthy automatic metric
from an unverified one.

% ============================================================
\section{The ConflictAgent Harness}
\label{sec:method}

\begin{figure}[t]
\centering
\includegraphics[width=0.92\linewidth]{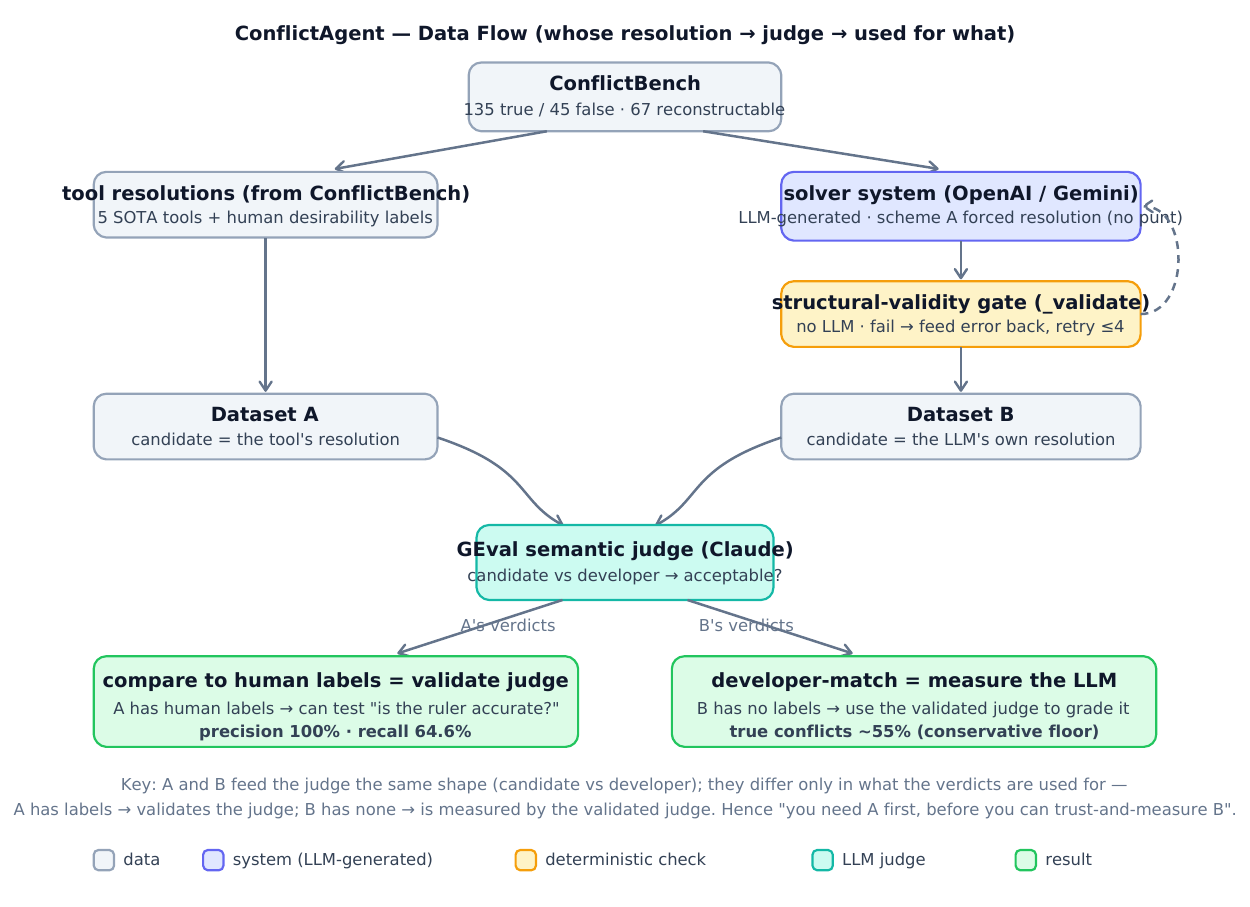}
\caption{The two-layer data flow. The solver (top) generates candidates without ground truth; the
evaluation layer (bottom) uses ground truth. Dataset~A grades the \emph{tools'} recorded outputs
(which carry human labels) to \emph{validate} the judge; Dataset~B then uses the validated judge to
\emph{measure} the LLMs. This is why the judge must be calibrated on~A before it is trusted on~B.}
\label{fig:dataflow}
\end{figure}

\subsection{Data}
\label{sec:data}
We use ConflictBench~\cite{shen2024conflictbench} as our data source. It contains 180 textual merge
scenarios sampled from 180 open-source Java projects; each scenario ships the \texttt{base},
\texttt{left}, \texttt{right}, and developer (\texttt{child}) versions, the outputs of the five merge
tools when available, and human labels including per-tool desirability judgments. ConflictBench
labels 136 of the 180 conflicts as true and 44 as false. One scenario (\texttt{orientdb@501dac79})
is a line-ending artifact that \texttt{git merge} resolves cleanly after newline normalization; we
treat it as a false conflict, giving an effective 135/45 split. This reclassification has no effect
on any reported number, since the scenario is already excluded upstream as a non-conflict.

\begin{figure}[t]
\centering
\includegraphics[width=0.95\linewidth]{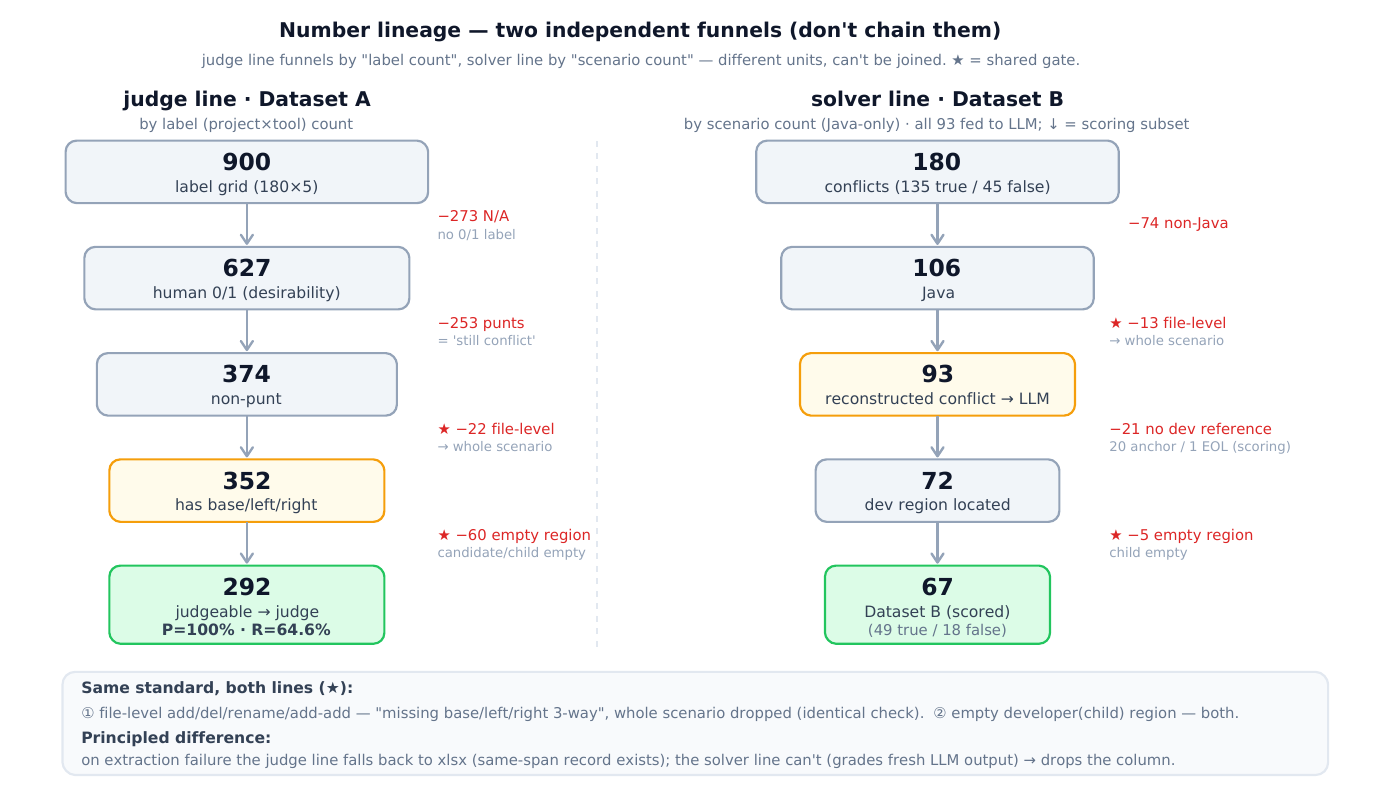}
\caption{Two independent number funnels that must \emph{not} be chained: the judge line (left) is
counted in labeled (scenario\,$\times$\,tool) pairs and ends at the $n{=}292$ meta-evaluation set;
the solver line (right) is counted in Java scenarios and ends at the 93 reconstructable conflicts fed
to each LLM. Shared structural gates are marked~$\star$.}
\label{fig:funnel}
\end{figure}

Because our structural validity check is Java-specific and requires a reconstructable in-file
three-way merge, our solver-evaluation scope is a subset of the full benchmark
(Fig.~\ref{fig:funnel}, right). Of the 180 scenarios, 106 are Java; of those, 93 are reconstructable
from complete base/left/right files. All 93 are fed to each LLM; the smaller numbers reported later
are \emph{scoring} subsets (cases we can reliably grade), not input samples.

\subsection{The solver agent (no ground truth)}
\label{sec:solver}
\begin{figure}[t]
\centering
\includegraphics[width=0.9\linewidth]{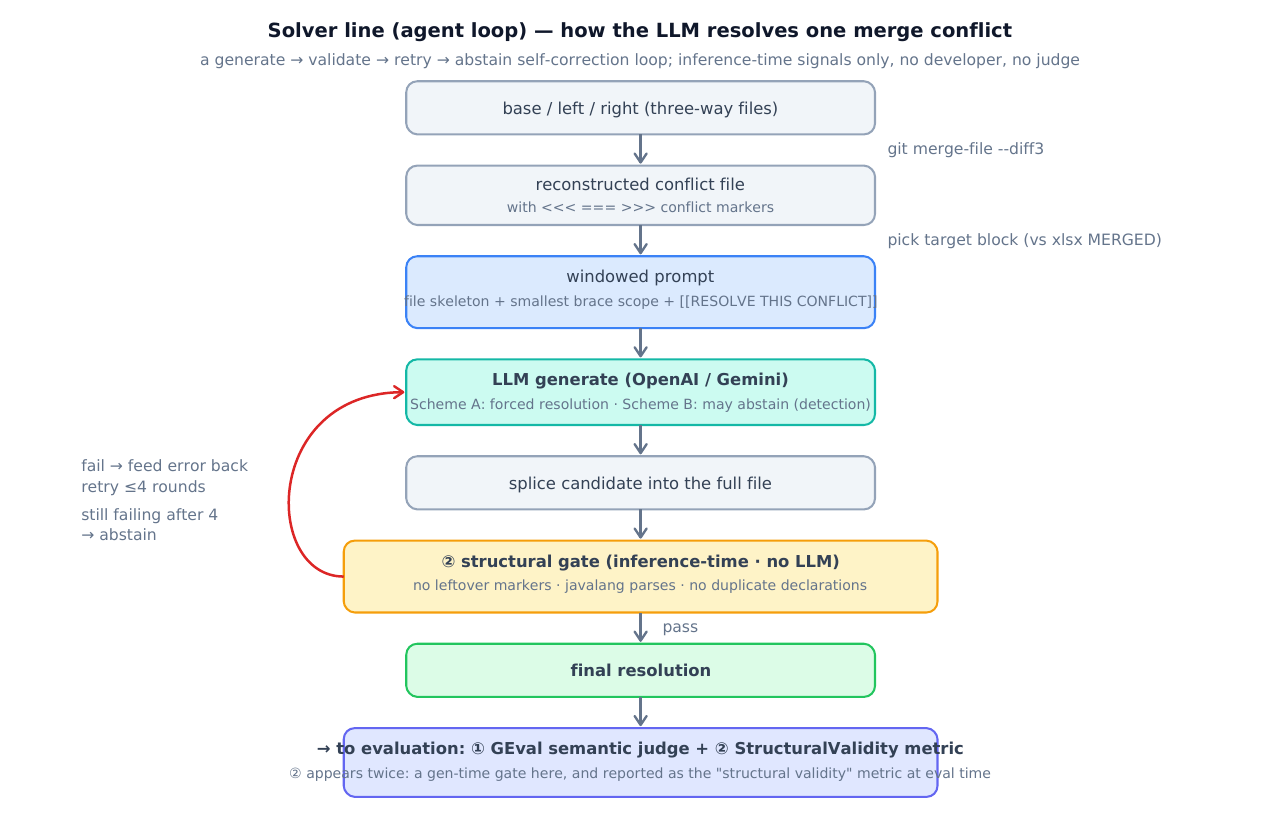}
\caption{The solver agent loop: reconstruct the diff3 file, window the prompt around the target
block, generate a resolution, splice it into the full file, and run the deterministic structural gate
(no leftover markers, parses via \texttt{javalang}, no duplicate declarations). Validation failures
feed the error back for up to four retries; persistent failure abstains. The loop uses inference-time
signals only---no developer answer, no judge.}
\label{fig:solver}
\end{figure}

The solver layer (Fig.~\ref{fig:solver}) turns one scenario into one candidate resolution through a
generate--validate--retry loop:

\begin{enumerate}
\item Reconstruct a \texttt{diff3} conflict file from base/left/right with
\texttt{git merge-file -{}-diff3}.
\item Select the target conflict block by matching content from ConflictBench's annotated merged
snippet against the reconstructed blocks (files may contain several blocks).
\item Build a \emph{windowed} prompt: the package/import/type skeleton plus the smallest complete
brace scope enclosing the target block, with the rest of the file elided. Files below a size
threshold are shown whole. Windowing affects only what the model sees; validation and splicing always
use the full reconstructed file.
\item Ask the solver to replace \emph{only} the tagged conflict region.
\item Splice the candidate back into the full file.
\item Validate using inference-time signals only: no leftover conflict markers, the spliced file
parses as Java via \texttt{javalang}~\cite{javalang}, and no duplicate declarations (which catch
over-scoped output).
\item Retry up to four times when validation fails, feeding the validator error back to the model;
if it still fails, abstain.
\end{enumerate}

Two prompt schemes are kept as methodology but never pooled. \textbf{Scheme~A} (the primary, scored
scheme) forces the model to produce a resolution plus a self-reported strategy and confidence.
\textbf{Scheme~B} is a detection/robustness variant that first lets the model declare a
\texttt{TRUE\_CONFLICT} and abstain; it is reported as a separate capability, not folded into the
Scheme-A headline. The exact prompts are in the released harness.

We evaluate two solvers of different vendors, \texttt{gpt-5.4-2026-03-05} (OpenAI) and
\texttt{gemini-3.5-flash} (Google), both at temperature~0 for reproducibility.

\subsection{The evaluation suite (ground truth allowed)}
\label{sec:eval}
Evaluation runs after the agent finishes and is the only place ground truth enters. It is built on
DeepEval~\cite{deepeval} and comprises two orthogonal metrics (Fig.~\ref{fig:judge}):

\begin{figure}[t]
\centering
\includegraphics[width=0.95\linewidth]{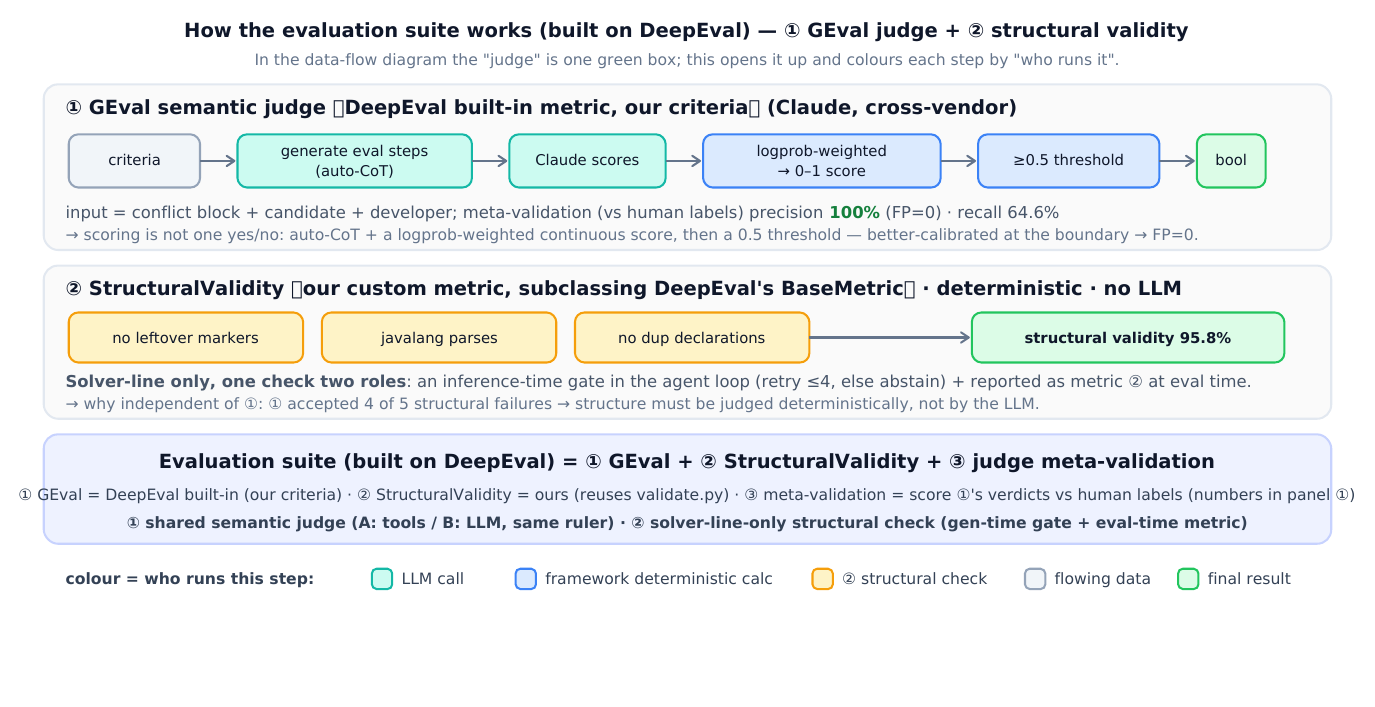}
\caption{Inside the evaluation suite. Metric~\ding{192} (developer-match) is a G-Eval LLM judge that
generates chain-of-thought evaluation steps and a probability-weighted score before thresholding.
Metric~\ding{193} (structural validity) is a deterministic, non-LLM check. Meta-validation~\ding{194}
scores \ding{192}'s verdicts against human labels. Colors indicate which component runs each step.}
\label{fig:judge}
\end{figure}

\paragraph{① Developer-match (LLM judge).}
A G-Eval~\cite{liu2023geval} metric (\emph{ResolutionAcceptability}) sees the diff3 conflict block,
the candidate resolution, and the developer's actual resolution, and decides whether the candidate
is an acceptable \emph{semantic} match (not a verbatim match, which would be far too strict). The
judge is \texttt{claude-sonnet-4-6}---deliberately a different vendor from both solvers to avoid
self-preference---at temperature~0.

\paragraph{② Structural validity (deterministic, no LLM).}
A candidate is structurally valid if it has no leftover markers, parses as Java, and introduces no
duplicate declarations. This reuses the same three checks the agent applies at generation time, but
here it is reported as an independent metric. Critically, ② does \emph{not} gate ①: we report them
separately precisely so we can ask whether the LLM judge is reliable about structural correctness
(Sec.~\ref{sec:results-struct}).

\paragraph{Baselines.} Because a strong first-round solver usually emits syntactically valid output,
an early single-shot baseline carried little signal and was dropped. We instead compare against four
interpretable trivial baselines---\texttt{pick-left}, \texttt{pick-right}, \texttt{pick-longer}, and
\texttt{union}---and against the five ConflictBench tools, all scored by the \emph{same} ① judge.
A secondary \emph{standalone-valid} judge (is the candidate reasonable from base/left/right alone,
without the developer answer?) is retained but is meaningful only for false conflicts, which have an
objective mechanical merge; it is deliberately \emph{excluded} from the headline because it has no
ground truth to validate against.

% ============================================================
\section{Calibrating the Judge}
\label{sec:judge}

Before using the ① judge to grade LLM resolutions, we validate it against ConflictBench's human
labels (a \emph{meta-evaluation}). The idea (Fig.~\ref{fig:dataflow}) is that the tool outputs
already carry human desirability labels, so we can ask the judge to grade those same tool outputs and
compare its verdicts to the humans'.

\paragraph{Population.} The meta-evaluation funnel (Fig.~\ref{fig:funnel}, left) is measured in
labeled (scenario\,$\times$\,tool) pairs, not scenarios. It starts from the $180\times5=900$ label
grid and removes, in order: 273 cells with no 0/1 desirability label; 253 tool punts (the tool left
the conflict unresolved---a detection event, not a resolution); 22 file-level operations
(add/delete/rename, for which there is no reconstructable in-file three-way merge---the same
structural gate the solver applies at $106\to93$); and 60 empty regions (candidate or developer empty
after extraction). This leaves \textbf{$n{=}292$ judgeable cases}.

\paragraph{Result.} Against the human labels, at the default acceptance threshold 0.5:

\begin{table}[t]
\centering
\caption{Judge meta-evaluation against human desirability labels ($n{=}292$,
judge \texttt{claude-sonnet-4-6}, threshold 0.5). The judge has zero false accepts.}
\label{tab:judge}
\begin{tabular}{lr}
\toprule
Metric & Value \\
\midrule
Accuracy  & 78.1\% \\
\textbf{Precision} & \textbf{100.0\%} \\
Recall    & 64.6\% \\
\midrule
True positives (TP)  & 117 \\
False positives (FP) & \textbf{0} \\
True negatives (TN)  & 111 \\
False negatives (FN) & 64 \\
\bottomrule
\end{tabular}
\end{table}

Table~\ref{tab:judge} reports the headline: \textbf{precision 100\%} (zero false accepts) at recall
64.6\%. The interpretation drives the rest of the paper. A false accept---calling a bad resolution
acceptable---is the fatal error for a judge, because it would inflate every downstream rate; the
judge makes zero of these on this set. The cost is recall: the judge under-credits some acceptable
alternatives (it is stricter than the median human annotator). Consequently every solver or tool
rate produced by this judge is a \textbf{conservative lower bound}, and an \texttt{ACCEPTABLE}
verdict can be trusted.

\paragraph{Why high precision is not an artifact of clean inputs.} A residual minority of label
pairs record the candidate and developer at inconsistent scopes (e.g., a 2-line developer window
versus a 24-line tool block). Such a mismatch can only cause the judge to \emph{reject} a valid
resolution---depressing recall---never to produce a false accept. Precision therefore stays at 100\%
regardless, and the conservative-lower-bound conclusion is stable.

\paragraph{Standalone-valid (secondary).} On a separate 40-item human-labeled blind sample, the
standalone-valid judge scores accuracy 85\%, precision 88.2\%, recall 93.8\% on the 20 false
conflicts (TP15/FP2/TN2/FN1). Only the false-conflict figure is a substantive correctness number,
because a true conflict has no context-free correct answer; this metric is retained as supplementary
and kept out of the headline.

% ============================================================
\section{Results}
\label{sec:results}

We organize results around Q2. Section~\ref{sec:results-dev} reports developer-match rates,
Section~\ref{sec:results-tools} compares against traditional tools under a coverage-fair convention,
and Section~\ref{sec:results-struct} reports structural validity and the judge-reliability finding.

\paragraph{Scored population.} Extending the Scheme-A branch of Fig.~\ref{fig:funnel}, the 72
comparable scenarios lose 5 with an empty developer region, giving \textbf{67 Dataset~B scenarios
(49 true / 18 false)}. Fed to both providers this yields 132 records (OpenAI 67, Gemini 65---Gemini
produced two empty resolutions that cannot be graded), i.e.\ 96 true and 36 false provider-level
records.

\subsection{How often do LLMs match the developer?}
\label{sec:results-dev}

\begin{table}[t]
\centering
\caption{① Developer-match rate (validated G-Eval judge, threshold 0.5), by conflict type and
provider. The headline quotes the conservative floor $\approx$55\% (OpenAI on true conflicts).}
\label{tab:devmatch}
\begin{tabular}{lccc}
\toprule
Conflict type & OpenAI & Gemini & Pooled \\
\midrule
\textbf{True} & 27/49 = 55.1\% & 29/47 = 61.7\% & 56/96 = 58.3\% \\
False         & 12/18 = 66.7\% & 11/18 = 61.1\% & 23/36 = 63.9\% \\
\bottomrule
\end{tabular}
\end{table}

Table~\ref{tab:devmatch} shows the ① developer-match results. On true conflicts---the hard case
requiring judgment---the two solvers match the developer's own resolution 55.1\% (OpenAI) and 61.7\%
(Gemini) of the time. We quote the \textbf{conservative floor, $\approx$55\%} (the lower provider),
rather than the pooled 58.3\%, for two reasons: the providers resolve different subsets so a pooled
average is not meaningful, and reporting a floor is consistent with the judge being a strict lower
bound. Because the judge counts only developer-\emph{matching} resolutions---not resolutions that are
valid but simply different from what the developer chose---this $\approx$55\% is a lower bound in a
second sense as well.

\subsection{LLMs versus traditional tools: a coverage-fair comparison}
\label{sec:results-tools}

The central comparison pits the LLM against the five ConflictBench tools \emph{on the same judged
scale}. All six are scored by the same ① judge on the 49 true conflicts of the reconstructable-Java
overlap. The fair convention when one side may abstain is to \emph{fix the denominator by the side
that always answers} (the LLM under Scheme~A) and count the other side's abstention as a miss rather
than shrinking the denominator---otherwise a tool's biggest weakness, abstaining, would be hidden.

\begin{table}[t]
\centering
\caption{① LLM vs.\ five traditional tools on true conflicts ($n{=}49$), same judge, coverage-fair.
``Among-resolved'' scores only cases where the tool produced a resolution; ``Overall'' counts a punt
as a miss (the coverage-fair headline). ``Punt'' is the count of abstentions.}
\label{tab:tools}
\begin{tabular}{lccc}
\toprule
Resolver & Among-resolved & \textbf{Overall} & Punt \\
\midrule
LLM Gemini    & 29/47 = 61.7\% & \textbf{29/49 = 59.2\%} & 0 \\
LLM OpenAI    & 27/49 = 55.1\% & \textbf{27/49 = 55.1\%} & 0 \\
AutoMerge     & 18/38 = 47.4\% & 18/49 = 36.7\%          & 10 \\
JDime         & 17/31 = 54.8\% & 17/49 = 34.7\%          & 16 \\
IntelliMerge  & 13/27 = 48.1\% & 13/49 = 26.5\%          & 22 \\
FSTMerge      & \phantom{0}6/21 = 28.6\% & \phantom{0}6/49 = 12.2\% & 18 \\
KDiff3        & \phantom{0}2/4\phantom{0} = 50.0\% & \phantom{0}2/49 = \phantom{0}4.1\% & 45 \\
\bottomrule
\end{tabular}
\end{table}

Table~\ref{tab:tools} contains the paper's main empirical result. \emph{Among resolved cases}, the
LLM and the strongest tool are close (Gemini 61.7\% vs.\ JDime 54.8\%): when a structured tool does
fire, its accuracy is competitive. The gap opens entirely on \textbf{coverage}. The tools abstain
heavily---KDiff3 on 45 of 49 conflicts, IntelliMerge on 22, JDime on 16, AutoMerge on 10---while the
LLM under Scheme~A abstains on none. Under the coverage-fair overall convention the LLMs (55--59\%)
clear the strongest tool (AutoMerge, 36.7\%) by roughly 18--22 points.

The qualitative conclusion is deliberately narrow: the LLM's advantage is \emph{not} ``substantially
more accurate'' but ``still produces a gradeable resolution where structured tools give up.'' That is
the applicability claim, and it is the honest reading of Table~\ref{tab:tools}.

\subsection{Structural validity, and when \emph{not} to trust an LLM judge}
\label{sec:results-struct}

\begin{table}[t]
\centering
\caption{② Structural validity (deterministic, no LLM), by conflict type.}
\label{tab:struct}
\begin{tabular}{lcc}
\toprule
Conflict type & Valid & Rate \\
\midrule
True  & 92/96 & \textbf{95.8\%} \\
False & 35/36 & 97.2\% \\
\bottomrule
\end{tabular}
\end{table}

Table~\ref{tab:struct} shows structural validity is high: 95.8\% on true conflicts. The 4.2\% that
fail are all retries-exhausted cases that never reached a valid resolution within the four-round
budget---three \texttt{javalang} parse failures and two duplicate-declaration cases.

The important finding is about the relationship between the two metrics. Of the five resolutions that
fail the deterministic structural check ②, \textbf{four were accepted by the LLM judge ①}: the LLM
judge waved through code that does not parse. This is concrete evidence that structural correctness
must be evaluated by a deterministic checker, \emph{not} delegated to an LLM, and it is why we keep ②
independent of ① rather than letting the judge subsume it. Knowing when \emph{not} to use an LLM is
part of using one well.

% ============================================================
\section{Case Studies}
\label{sec:cases}

To make the aggregate numbers concrete, we walk through six representative scenarios (drawn from the
saved evaluation records) that each illustrate one design decision behind the harness. A caveat on
reading them: the per-case verdicts in Table~\ref{tab:cases} are saved records used to \emph{explain}
the evaluation design (why we need trivial baselines, why standalone-valid and developer-match answer
different questions, why extraction is guarded); these design points are judge-agnostic. The
\emph{reported} rates are the aggregate G-Eval numbers in Section~\ref{sec:results}; individual
per-case entries here are anchors, not the headline figures.

\begin{table*}[t]
\centering
\caption{Six case studies, each illustrating one evaluation-design point. ``T/F'' is the
ConflictBench human label (true/false conflict). \emph{dev-match} = judged an acceptable match to the
developer's resolution; \emph{standalone} = judged reasonable from base/left/right alone;
\emph{final-valid} = passed structural validation after splicing. ``--'' = not applicable (e.g., an
excluded scenario or a punt).}
\label{tab:cases}
\footnotesize
\begin{tabular}{@{}l c p{5.9cm} p{4.3cm}@{}}
\toprule
Scenario & T/F & Observation (as recorded) & Design point it anchors \\
\midrule
\texttt{error-prone@6f83c083} & T & Both LLMs match the developer; all four trivial baselines miss
(dev-match false). & LLMs beat heuristics on a genuine conflict. \\[2pt]
\texttt{dubbo@b7b34b6c} & T & Both LLMs match the developer; a side-pick can look plausible in
isolation (standalone true) yet miss the developer's intent (dev-match false). & Why we replaced the
single-shot baseline with trivial baselines. \\[2pt]
\texttt{mybatis-3@3502f7ce} & T & Both LLMs miss the developer; the \texttt{pick-right} baseline
matches it. & Baselines are necessary---LLMs can over-merge. \\[2pt]
\texttt{RxJava@45c9dc85} & F & Candidates look reasonable standalone but do not match the developer;
OpenAI additionally fails structural validation after splice. & Standalone-valid, developer-match,
and structural-validity are three different questions. \\[2pt]
\texttt{proxyee-down@1d9d7f71} & F & Agent produces and validates a resolution, but context anchors are
not unique, so the developer region is not extractable. & Guarded extraction: why denominators are
smaller than 93 (a data-integrity guard, not a model failure). \\[2pt]
\texttt{Matisse@93d0051c} & T & Under Scheme~B both solvers punt (declare \texttt{TRUE\_CONFLICT}); the
human label agrees. & Detection: a punt is a correct abstention event, not a failed resolution. \\
\bottomrule
\end{tabular}
\end{table*}

\paragraph{The value case (\texttt{error-prone}, \texttt{dubbo}).} These are true conflicts where every
trivial baseline misses the developer's resolution but both LLMs match it. They are the concrete form
of the headline: on genuine conflicts an LLM can recover the developer's intent where side-picking and
union heuristics cannot.

\paragraph{The anti-overclaim case (\texttt{mybatis-3}).} Here the developer simply took one side, and
the \texttt{pick-right} baseline captures that while both LLMs produce a reasonable-but-different
merge. This is exactly why we report trivial baselines alongside the LLM: without them, an LLM that
``looks smart'' could be losing to a one-line heuristic on the cases that matter.

\paragraph{Three questions, not one (\texttt{RxJava}).} Several candidates are reasonable in isolation
(\emph{standalone} true) yet are not the developer's answer (\emph{dev-match} false), and OpenAI's
candidate additionally fails structural validation after being spliced back
(\emph{final-valid} false). This single scenario shows why the suite keeps developer-match, the
secondary standalone-valid check, and the deterministic structural metric as separate axes.

\paragraph{Guarded ground truth (\texttt{proxyee-down}).} The agent resolves the conflict, but the
evaluation layer refuses to guess the developer's corresponding region when context anchors are not
unique, so the scenario is excluded from the developer-match denominator. This is the mechanism behind
the conservative $93\!\to\!72$ scoring funnel: the guard abstains rather than risk a wrong label.

\paragraph{Abstention as a capability (\texttt{Matisse}).} Under the Scheme-B detection variant, both
solvers decline to auto-resolve and the human label agrees that the conflict is genuinely true. A punt
is scored as a correct detection event, not as a failed resolution---the same abstention that, for the
traditional tools, we count as a coverage miss in Table~\ref{tab:tools}.

% ============================================================
\section{Discussion}
\label{sec:discussion}

\paragraph{Coverage is the story, not accuracy.} The temptation with LLM results is to headline a
single accuracy number. Our comparison resists that: among resolved cases the LLM barely edges the
best tool. What changes the picture is that structured tools decline to answer most of the time. In
practice a developer facing a conflict benefits more from a gradeable proposal on 100\% of conflicts
than from a marginally better proposal on the 10--80\% a tool is willing to touch.

\paragraph{Everything is a lower bound.} Three design choices all push our numbers down rather than
up: the judge is strict (100\% precision, 64.6\% recall), so it under-credits acceptable
alternatives; we count only developer-matching resolutions, excluding valid-but-different ones; and
we quote the lower-provider floor rather than a pooled average. We prefer to under-claim with a
defensible protocol than to over-claim with an unvalidated one.

\paragraph{Calibrate the ruler before you measure with it.} The methodological point generalizes
beyond merge conflicts. An LLM-as-judge is a measurement instrument, and an uncalibrated instrument
produces uninterpretable numbers. Validating the judge against human labels \emph{first}---and
reporting precision, not just accuracy---is what lets us attach a direction (conservative) to every
result. We would encourage other LLM-evaluation work to report the judge's own confusion matrix
against ground truth before reporting anything the judge measures.

% ============================================================
\section{Threats to Validity}
\label{sec:threats}

\paragraph{Construct validity.} The developer-match metric equates ``good resolution'' with
``matches what the developer committed.'' A different-but-valid resolution is scored as a miss; our
numbers are lower bounds accordingly. LLM outputs also lack independent human labels, so residual
judge bias cannot be fully excluded---mitigated, but not eliminated, by using a judge from a
different vendor than both solvers.

\paragraph{Internal validity.} Developer resolutions are extracted from the \texttt{child} file using
textual context anchors. When anchors are not unique the scenario is \emph{excluded} rather than
guessed; 20 Scheme-A scenarios are excluded this way, splitting into four causes (boundary edges 8,
duplicated boilerplate context 6, adjacent-block markers in multi-block files 4, and developer
rewrites that erased the anchor 2). The guard excludes rather than guesses, so the denominator is a
conservative subset, not a biased one. Separately, \texttt{javalang} checks syntax, not full
compilation, and for multi-block files still carrying other blocks' markers the structural check
degrades to a marker-only check.

\paragraph{External validity.} The study is Java-only and, for the solver evaluation, restricted to
the 93 reconstructable scenarios (67 in Dataset~B); results may not transfer to other languages or to
conflicts outside ConflictBench's sampling. The tool baselines use ConflictBench's recorded outputs
rather than locally re-run tools. Finally, LLMs evolve quickly; the specific rates are tied to the two
2026-era models evaluated, though the methodology is model-agnostic.

% ============================================================
\section{Conclusion}
\label{sec:conclusion}

We studied whether modern LLMs can resolve real Java merge conflicts, and---just as importantly---how
to measure that credibly. By calibrating an LLM-as-judge against human labels before using it
(precision 100\%, recall 64.6\%, $n{=}292$), we turned every downstream rate into a conservative
lower bound. Under that judge, LLMs match the developer's own resolution on $\approx$55\% of true
conflicts and, under a coverage-fair comparison, beat the strongest traditional merge tool by 18--22
points---an advantage that comes from coverage, since structured tools abstain on much of the
workload while the LLM does not. We also showed that structural correctness must stay a deterministic
check: the LLM judge accepted four of five resolutions that fail to parse. The harness is released as
an open engineering artifact to support replication and extension to other languages and models.

% ============================================================
\section*{Data and Code Availability}
The evaluation harness is available at \url{https://github.com/UBOWENVT/ConflictAgent}. It builds on
ConflictBench~\cite{shen2024conflictbench}, which supplies the data, developer resolutions, human
labels, and tool baselines.

% ------------------------------------------------------------
% Bibliography inlined (no external .bib / bibtex needed).
% A matching refs.bib is kept alongside for record-keeping; to
% switch back to BibTeX, replace this block with:
%   \bibliographystyle{elsarticle-num}\bibliography{refs}
% ------------------------------------------------------------

\end{document}